\documentclass[sigconf]{acmart}

\usepackage{booktabs} 
\usepackage{mdframed} 
\usepackage{paralist}
\usepackage{graphicx} 
\usepackage{enumitem}
\usepackage{balance}

\setcopyright{rightsretained}

\acmConference[SCAI'17]{ICTIR'17 Workshop on Search-Oriented Conversational AI (SCAI'2017)}{October 1, 2017}{Amsterdam, Netherlands}
\acmYear{2017}
\copyrightyear{2017}

\begin{document}

\title{Conversational Exploratory Search via Interactive Storytelling}
\subtitle{Position Paper}

\author{Svitlana Vakulenko}
\affiliation{%
  \institution{Vienna University of
Economics and Business}
  \city{Vienna} 
  \country{Austria}
}
\email{svitlana.vakulenko@wu.ac.at}

\author{Ilya Markov}
\affiliation{%
  \institution{University of Amsterdam}
  \city{Amsterdam} 
  \country{The Netherlands}}
\email{i.markov@uva.nl}

\author{Maarten de Rijke}
\affiliation{%
  \institution{University of Amsterdam}
  \city{Amsterdam} 
  \country{The Netherlands}}
\email{derijke@uva.nl}

\renewcommand{\shorttitle}{Conversational Exploratory Search}

\settopmatter{printacmref=false}

\begin{abstract}

Conversational interfaces are likely to become more efficient, intuitive and engaging way for human-computer interaction than today's text or touch-based interfaces. Current research efforts concerning conversational interfaces focus primarily on question answering functionality, thereby neglecting support for search activities beyond targeted information lookup. Users engage in exploratory search when they are unfamiliar with the domain of their goal, unsure about the ways to achieve their goals, or unsure about their goals in the first place. Exploratory search is often supported by approaches from information visualization. However, such approaches cannot be directly translated to the setting of conversational search.

In this paper we investigate the affordances of interactive storytelling as a tool to enable exploratory search within the framework of a conversational interface. Interactive storytelling provides a way to navigate a document collection in the pace and order a user prefers. In our vision, interactive storytelling is to be coupled with a dialogue-based system that provides verbal explanations and responsive design. We discuss challenges and sketch the research agenda required to put this vision into life.

\end{abstract}


%
%







\maketitle


\section{Introduction}
\label{sec:intro}

Exploratory search systems provide guidance for users who are exploring unfamiliar information landscapes~\cite{Marchionini:2006:Exploratory,DBLP:series/synthesis/2009White}. \citet{DBLP:series/synthesis/2009White} differentiate two main activities within the exploratory search paradigm: \emph{exploratory browsing} and \emph{focused searching}. Exploratory browsing is an initial step that provides necessary domain understanding required for focused searching activities. It is related to \citet{DBLP:conf/chiir/RadlinskiC17}'s \emph{system revealment} property: ``The system reveals to the user its capabilities and corpus, building the user's expectations of what it can and cannot do.''





Lately, conversational agents and conversational search systems are becoming increasingly popular \citep{DBLP:journals/corr/Thomas17}.
So far, however, such systems mainly focus on question answering and simple search tasks, those that are to a large extent solved by web search engines.
We argue that conversational agents and search systems should also support exploratory search.
While exploratory search is a challenging task in itself,
\textit{conversational exploratory search} raises unique research and practical issues, which we discuss in this position paper.

In particular, we argue that the core of conversational exploratory search is \textit{interactive storytelling},
where the document collection underlying a conversational search system is first converted into a set of stories
and then a user interactively navigates within a story and between stories by means of a dialogue with the system.

There have been several recent position statements on conversational agents and search. One, by \citet{DBLP:conf/chiir/RadlinskiC17}, focuses on a theoretical model of conversational search systems. Another, by \citet{DBLP:journals/corr/KiselevaR17}, focuses on evaluation. In contrast, we focus on solution strategies for a specific conversational search scenario, viz.\ exploratory search.

Below, we first provide motivating examples in Section~\ref{sec:example}.
In Section~\ref{sec:system} we present our view of a conversational exploratory search system.
The research agenda associated with this system is presented in Section~\ref{sec:agenda}.
The paper is concluded in Section~\ref{sec:conclusions}.

\section{Motivating Examples}
\label{sec:example}

The literature is full of arguments motivating computational support for exploratory search~\citep{white-interactions-2016}. Exploratory search is an important enabler for educational purposes that aim to broaden the knowledge of a user and understanding of the domain by enhancing learning processes.
Serendipitous discoveries are very important in less structured and content rich domains such as music, videos, design etc., where users often look for inspiration, surprises and novel ideas~\citep{DBLP:conf/wsdm/ZhangSQJ12}.
Furthermore, the potential benefits of conversational exploratory search for e-commerce applications should not be underestimated. In particular, it can be combined with personal recommendations and persuasion techniques for marketing purposes~\citep{creativity}.

For us, one of the main motivations behind conversational exploratory search comes from the results of analyzing the conversation log of a chatbot demo that some of the authors were involved with~\citep{DBLP:journals/corr/NeumaierSV17}.\footnote{\url{https://m.me/OpenDataAssistant}} This chatbot demo exposes search functionality
over an aggregated open data repository~\cite{DBLP:conf/www/NeumaierUP17} via a conversational interface.
Manual inspection of the conversation log of the digital assistant revealed that the majority of users experience difficulties formulating adequate queries to the system, i.e., queries that return any matches. 
This effect is, to a large extent, due to a misconception of the underlying collection of documents, which can potentially be retrieved using a search system.

The observation of a user's mistaken internal representation of a document collection is not new. The information seeking literature is full of examples to this effect and the information retrieval community has proposed a range of technological solutions to help address such mismatches, ranging from algorithms that help recover from possibly empty search engine result pages using query suggestions and rewrites~\citep{li-investigating-2017} to information visualization techniques to help steer users in possibly useful regions of a document collection~\citep{hearst-search-2009}.

However, visualization approaches may vary significantly and  be hard to understand without an animated explanation in natural language or even specialized training. While such methods may be effective in traditional keyboard or touch-based exploratory search scenarios, by and large they are inappropriate to support exploratory search in a conversational setting on mobile devices. Instead, we argue that an approach based on interactive storytelling is called for to support conversational exploratory search.

\section{Conversational Exploratory Search}
\label{sec:system}

\begin{figure*}
\includegraphics[clip,trim=0mm 15mm 0mm 0mm,width=\textwidth]{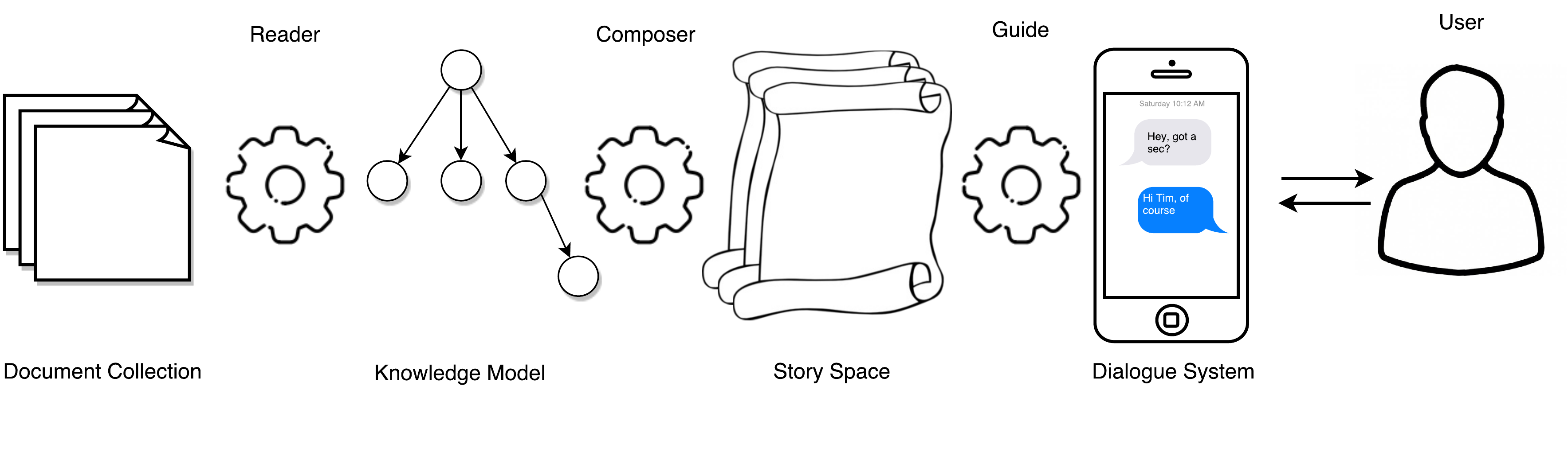}
\caption{Communicating knowledge via an interactive storytelling process.}
\label{fig:architecture}
\end{figure*}

Our view of a conversational exploratory search system is represented in Figure~\ref{fig:architecture}. It has a number of key components: Document Collection, Knowledge Model, Story Space, Dialog System and User. These components are connected through the Reader, Composer, and Guide modules. The interplay of the system components and modules happens at different stages.


\subsubsection*{Knowledge Representation} Knowledge representation consists of the \textit{Reader} module that extracts concepts and relations from the Document Collection and embeds them into a single Knowledge Model.
The Knowledge Model integrates different elements (words, concepts or entities) and describes relations between them.
The knowledge can be explicitly modeled by means of a taxonomy or ontology (knowledge graph) but it can also be embedded into a latent (hidden) structure.

\subsubsection*{Story Generation} Story generation consists of the \textit{Composer} module that is able to generate stories by combining elements of the Knowledge Model.
To create a story, the Composer has to select elements (characters, words, facts, concepts, relations), choose their ordering, arrange selected elements in time and/or space.
The set of all possible stories constitutes the Story Space.

\subsubsection*{Interactive Storytelling} Interactive storytelling consists of the \textit{Guide} module that helps the User to navigate through the Document Collection via the Story Space. The Guide can change the current position within a single story or traverse the space across different stories.
Interactive storytelling integrates the Dialogue System to communicate a story to the User and to receive an input from the User. Supporting such a conversation with the User requires natural language (utterance) generation and understanding. Note that the input/output modalities do not have to be restricted to text and speech only and may include images, videos, interactive visualization, virtual reality interactions, etc.






\medskip
We also argue that a conversational exploratory search system should support the following types of the user-system interactions:
\begin{itemize}

\item \textit{Navigation Control} -- a user chooses a direction (branch) for exploration and is also able to influence and change the current direction of the narrative at any point in time;


\item \textit{Feedback} -- a user may provide feedback to the system (positive, neutral, negative) that may help to correct and steer the direction of the story that shall maximize the user satisfaction with the system;

\item \textit{Question} -- a user may pose questions to the system, e.g., a request for a definition, look up query, etc.
\end{itemize}

A sample dialogue with interactions of all these types is provided in Figure~\ref{fig:dialogue}. In this example the dialogue agent provides concise natural language descriptions of the information space structure, suggestions for possible exploration directions, and further support and guidance along the chosen direction for exploration.

\begin{figure}
\begin{mdframed}

\begin{description}[leftmargin=14pt]
\item[System:] I am here to help you explore open data! This collection includes more than 18,000 open datasets from countries such as Italy, Ireland, Mexico, Austria, etc.
\item[User:] What data do you have about Austria?
\item[System:] Jobs, healthcare, population statistics, education, environment... Are you interested in anything from the above?
\item[User:] What else?
\item[System:] Traffic data, sports, agriculture, tourism.
\item[User:] Traffic sounds interesting! What kind of data is there?
\item[System:] There is data for different cities in Austria: Vienna, Linz, Salzburg, Innsbruck.
\item[User:] ok
\item[System:] For example, pedestrian zones in Vienna, bridges locations, Electronic timetable information for Linz.
\item[User:] Great! Show me something else.
\item[System:] There are 96 datasets about Education and Research in Austria. One of them is a List of all driving schools in Burgenland.
\item[User:] What is Burgenland?
\item[System:] Burgenland is the easternmost and least populous state of Austria [Wikipedia].
\end{description}
\end{mdframed}

\caption{Sample dialog for exploratory search based on the Open Data Assistant chatbot use case and the faceted-search interface of the Austrian Government Open Data portal \url{https://www.data.gv.at}.}
\label{fig:dialogue}
\end{figure}

\section{Research Agenda}
\label{sec:agenda}


We identify the following research questions with respect to the components and interaction types described in the previous section:
\begin{description}
\item[\textbf{RQ1}.] \textit{Reader}: How to model the information space structure, represent documents and relations between them for the purpose of story generation? 

\item[\textbf{RQ2}.] \textit{Composer}: How to generate a coherent narrative (story) that efficiently describes a knowledge model?

\item[\textbf{RQ3}.] \textit{Guide}: How to efficiently traverse/navigate a story space?

\item[\textbf{RQ4}.] \textit{Dialogue system}: How to provide support for the following three types of the system-user interactions:

\item[\textbf{RQ4.1}.] \textit{Storytelling}: How to communicate a story to a user?

\item[\textbf{RQ4.2}.] \textit{Question generation}: How to verify user understanding, satisfaction and preferences? 

\item[\textbf{RQ4.3}.] \textit{Response analysis}: How to interpret and correctly react to natural language utterances (or other signals), such as the ones expressing user satisfaction (feedback), communicating the desired directions for traversing the information space (navigation control), checking the terminology and asking other types of questions?
\end{description}

In the following, we organize these research questions into two subtasks, namely, story generation and interactive storytelling.


\subsection{Story Generation}

In the context of conversational AI we are primarily interested in developing an operational knowledge model (\textbf{RQ1}), i.e., the structure that the system can act upon, e.g., to answer questions or generate stories.
Story generation (\textbf{RQ2}) requires accomplishing the following three tasks:
\begin{inparaenum}[(1)]
  \item select elements of the knowledge model;
  \item choose an order in which to present these elements; and
  \item communicate the story to the user using the modalities available to the system, e.g., natural language and/or visualization (\textbf{RQ4.1}).
\end{inparaenum}

Computational narrative intelligence, the ability to craft, tell, understand and respond appropriately to narratives, is a core component of a strong AI system~\citep{DBLP:journals/corr/Riedl16,DBLP:phd/basesearch/Li15}. So far, it has mostly been developed with applications to fiction, in the context of computational creativity. We propose to put it to work for conversational exploratory search. To this end, we first recall some core concepts from the area and then sketch our ideas for putting it to work for conversational exploratory search.

\citet{DBLP:conf/acl/McIntyreL10} 
use genetic algorithms (GAs) to generate children stories from a corpus of fairy tales. They extract schemas from natural language texts using dependency parsing and co-reference resolution tools, then generate a single plot graph by merging these schemas. The plot graph constitutes the story space, where each path is a different story.
The algorithm then searches the story space for the best story candidates using a coherence function learned from training data~\citep{DBLP:journals/coling/BarzilayL08}. The produced stories are readable but short and uninformative, and can be considered as a proof-of-concept for the story generation approach.

\citet{DBLP:journals/corr/MartinAHSHR17} generate stories in natural language using two sequence-to-sequence recurrent neural networks (RNNs):
\begin{inparaenum}[(1)]
  \item event representations are extracted from text using dependency parsing, stemming and topic modeling;
  \item event2event RNN chains the extracted events together into stories;
  \item event2sentence RNN translates the generated story representation into natural language sentences.
\end{inparaenum}
This approach is applied to a corpus of movie plot summaries extracted from Wikipedia~\citep{DBLP:conf/acl/BammanOS13}. It is reported to achieve plausible and human-readable sentences.

\citet{DBLP:conf/naacl/HuangFMMADGHKBZ16} establish a new task of visual storytelling, in which the system is to generate a story in natural language given a sequence of images as an input. The baseline model for story generation is trained using sequence-to-sequence RNNs.

In our view, the work on algorithms for story generation is sufficiently mature so that it can be successfully used in the context of conversational exploratory search, especially to support dialogue management in conversational exploratory search, thereby offering the potential to address \textbf{RQ1}, \textbf{RQ2} and part of \textbf{RQ4}, namely, \textbf{RQ4.1}.

\subsection{Interactive Storytelling}
Conversational exploratory search is not a one way traffic. Hence, our perspective on using story generation for the purposes of conversational exploratory search needs to be complemented with conversational aspects.
Interactive storytelling is a conversation, in which a storyteller aims to convey a fraction of a knowledge model to a listener (\textbf{RQ4.1}), and the listener can actively influence the direction, flow and manner of the story being told (responsive by design, \textbf{RQ4.3}). Ability of the storyteller to ask questions and expose possible directions for exploration (\textbf{RQ4.2}) aims at encouraging listener's active engagement with the story and avoiding lengthy monologues in favor of a more balanced dialogue-based interaction with the content.

Approaches developed within the goal-oriented dialogue framework (Dialog State Tracking Challenge~\citep{DBLP:journals/dad/WilliamsRH16a}) are likely to be useful for dialogue management in the interactive storytelling settings as well. Within this framework the dialogue system is supported by a task-specific domain ontology. The ontology enumerates all concepts and attributes (slots) that a user can specify or request information for~\citep{DBLP:conf/acl/MrksicSWTY17}. 
The dialogue management model is trained to correctly classify user intents by matching user utterances to the elements in the domain ontology. It can also learn to use the distribution over intents to decide whether to execute an action or request a clarification from the user~\citep{DBLP:conf/acl/MrksicSWTY17}.


The results of the Dialog State Tracking Challenge show advantages of end-to-end dialog systems that employ discriminative models and embed a dialog directly as a sequence~\citep{DBLP:journals/dad/WilliamsRH16a}.
\citet{DBLP:journals/corr/BordesW16} show how to train such an end-to-end dialog system using the Memory Network architecture.
\citet{DBLP:conf/acl/DhingraLLGCAD17} use RNNs and reinforcement learning to train a dialogue system that can interactively retrieve items from a single table. 

\citet{DBLP:conf/acl/MrksicSWTY17} avoid the limitations of the exact word matching by loading pre-trained word vectors and composing them into intermediate representations to be able to scale to larger and more complex domains. They carry out an evaluation for a single domain (restaurants), which is described by an ontology with three attributes specifying the goal (information need) and eight attributes available for retrieval. While very promising for the task of conversational exploratory search, the question remains whether the proposed interactive storytelling approaches can scale up from the toy examples considered so far to support meaningful conversations using the full-sized knowledge graphs.

With the fraction of the knowledge model involved in communication getting bigger the major design challenges arise with respect to the balanced composition of the story space (\textbf{RQ2}) that allows efficient traversal and communication taking into account cognitive limitations of the human brain (\textbf{RQ3}). In addition, the ability to adopt useful shortcuts across the story space will reduce the traversal time and, thereby, improve the experience by avoiding linear search in favor of random access, when it is applicable (\textbf{RQ3}).

In addition to scale, another important challenge arises from the fact that interactive storytelling is different from a common conversational search task, where an agent tries to pin-point an item or an information subspace relevant to the user's query~\citep{DBLP:conf/chiir/RadlinskiC17}.
In this respect, interactive storytelling is hard to optimize, since there is no single correct answer.
We propose to measure the results of the interactive storytelling process with respect to:
\begin{inparaenum}[(1)]
  \item the learning outcomes, which constitute the fraction of the knowledge model gained on the listeners' side; and
  \item user satisfaction.
\end{inparaenum}
The datasets available for learning dialogue representations are currently limited to two types of tasks: general chit-chat and goal-oriented dialogues, such as restaurant reservation~\citep{DBLP:conf/sigir/KenterBGDRM17,DBLP:journals/corr/MillerFFLBBPW17}. 

There are a few new datasets of conversation transcripts covering more general search scenarios~\citep{trippas2016how,DBLP:journals/corr/Thomas17}, which focus primarily on analyzing different task complexity levels and user experience during the dialogue interactions. To the best of our knowledge, there is currently no publicly available dataset of conversation logs recorded for learning conversational exploratory browsing behavior and evaluation of successful knowledge transfer interactions. 

\section{Conclusions}
\label{sec:conclusions}

In this paper we introduced the idea of enabling conversational exploratory search by means of interactive storytelling. We presented our vision of such a system, its components and modules. We also outlined directions for future research towards development of the computational narrative intelligence, as an enabler of conversational AI, and its application in the exploratory search scenarios, which go beyond the discrete look-up requests towards continuous interaction sessions with the goal of knowledge transfer, that we refer to as interactive storytelling.

The insights gained in the fields of story generation and dialogue systems suggest that it is feasible to develop a computational model able to learn natural language generation and communication from crowd-sourced examples. We see our task in developing this idea further by adopting it in the context of exploratory search. To begin in this direction, the research community requires a collection of new datasets of dialogue interactions that can be used for evaluation of successful knowledge transfer. Next, evaluation of existing approaches to story generation and learning dialogue policies in this new settings will help to form the baselines for developing novel approaches.




\section*{Acknowledgments}

The work of Svitlana Vakulenko has received funding from the EU H2020 programme under the MSCA-RISE agreement 645751 ({RISE\_BPM}) and the Austrian Research Promotion Agency (FFG) under the project CommuniData (grant no. 855407). 
Ilya Markov and Maarten de Rijke were supported by
Ahold Delhaize,
Amsterdam Data Science,
the Bloomberg Research Grant program,
the Criteo Faculty Research Award program,
Elsevier,
the European Community's Seventh Framework Programme (FP7/2007-2013) under
grant agreement nr 312827 (VOX-Pol),
the Microsoft Research Ph.D.\ program,
the Netherlands Institute for Sound and Vision,
the Netherlands Organisation for Scientific Research (NWO)
under pro\-ject nrs
612.\-001.\-116, 
HOR-11-10, 
CI-14-25, 
652.\-002.\-001, 
612.\-001.\-551, 
652.\-001.\-003, 
and
Yandex.
All content represents the opinion of the authors, which is not necessarily shared or endorsed by their respective employers and/or sponsors.

\bibliographystyle{ACM-Reference-Format}
\bibliography{bibliography} 


\begin{thebibliography}{00}


\ifx \showCODEN    \undefined \def \showCODEN     #1{\unskip}     \fi
\ifx \showDOI      \undefined \def \showDOI       #1{#1}\fi
\ifx \showISBNx    \undefined \def \showISBNx     #1{\unskip}     \fi
\ifx \showISBNxiii \undefined \def \showISBNxiii  #1{\unskip}     \fi
\ifx \showISSN     \undefined \def \showISSN      #1{\unskip}     \fi
\ifx \showLCCN     \undefined \def \showLCCN      #1{\unskip}     \fi
\ifx \shownote     \undefined \def \shownote      #1{#1}          \fi
\ifx \showarticletitle \undefined \def \showarticletitle #1{#1}   \fi
\ifx \showURL      \undefined \def \showURL       {\relax}        \fi
\providecommand\bibfield[2]{#2}
\providecommand\bibinfo[2]{#2}
\providecommand\natexlab[1]{#1}
\providecommand\showeprint[2][]{arXiv:#2}

\bibitem[\protect\citeauthoryear{Bamman, O'Connor, and Smith}{Bamman
  et~al\mbox{.}}{2013}]%
        {DBLP:conf/acl/BammanOS13}
\bibfield{author}{\bibinfo{person}{David Bamman}, \bibinfo{person}{Brendan
  O'Connor}, {and} \bibinfo{person}{Noah~A. Smith}.}
  \bibinfo{year}{2013}\natexlab{}.
\newblock \showarticletitle{Learning Latent Personas of Film Characters}. In
  \bibinfo{booktitle}{{\em Proceedings of the 51st Annual Meeting of the
  Association for Computational Linguistics, {ACL} 2013, 4-9 August 2013,
  Sofia, Bulgaria, Volume 1: Long Papers}}. \bibinfo{pages}{352--361}.
\newblock
\showURL{%
\url{http://aclweb.org/anthology/P/P13/P13-1035.pdf}}


\bibitem[\protect\citeauthoryear{Barzilay and Lapata}{Barzilay and
  Lapata}{2008}]%
        {DBLP:journals/coling/BarzilayL08}
\bibfield{author}{\bibinfo{person}{Regina Barzilay} {and}
  \bibinfo{person}{Mirella Lapata}.} \bibinfo{year}{2008}\natexlab{}.
\newblock \showarticletitle{Modeling Local Coherence: An Entity-Based
  Approach}.
\newblock \bibinfo{journal}{{\em Computational Linguistics\/}}
  \bibinfo{volume}{34}, \bibinfo{number}{1} (\bibinfo{year}{2008}),
  \bibinfo{pages}{1--34}.
\newblock
\showDOI{%
\url{https://doi.org/10.1162/coli.2008.34.1.1}}


\bibitem[\protect\citeauthoryear{Bordes and Weston}{Bordes and Weston}{2016}]%
        {DBLP:journals/corr/BordesW16}
\bibfield{author}{\bibinfo{person}{Antoine Bordes} {and} \bibinfo{person}{Jason
  Weston}.} \bibinfo{year}{2016}\natexlab{}.
\newblock \showarticletitle{Learning End-to-End Goal-Oriented Dialog}.
\newblock \bibinfo{journal}{{\em CoRR\/}}  \bibinfo{volume}{abs/1605.07683}
  (\bibinfo{year}{2016}).
\newblock
\showURL{%
\url{http://arxiv.org/abs/1605.07683}}


\bibitem[\protect\citeauthoryear{Dhingra, Li, Li, Gao, Chen, Ahmed, and
  Deng}{Dhingra et~al\mbox{.}}{2017}]%
        {DBLP:conf/acl/DhingraLLGCAD17}
\bibfield{author}{\bibinfo{person}{Bhuwan Dhingra}, \bibinfo{person}{Lihong
  Li}, \bibinfo{person}{Xiujun Li}, \bibinfo{person}{Jianfeng Gao},
  \bibinfo{person}{Yun{-}Nung Chen}, \bibinfo{person}{Faisal Ahmed}, {and}
  \bibinfo{person}{Li Deng}.} \bibinfo{year}{2017}\natexlab{}.
\newblock \showarticletitle{Towards End-to-End Reinforcement Learning of
  Dialogue Agents for Information Access}. In \bibinfo{booktitle}{{\em
  Proceedings of the 55th Annual Meeting of the Association for Computational
  Linguistics, {ACL} 2017, Vancouver, Canada, July 30 - August 4, Volume 1:
  Long Papers}}. \bibinfo{pages}{484--495}.
\newblock
\showDOI{%
\url{https://doi.org/10.18653/v1/P17-1045}}


\bibitem[\protect\citeauthoryear{Hearst}{Hearst}{2009}]%
        {hearst-search-2009}
\bibfield{author}{\bibinfo{person}{Marti~A. Hearst}.}
  \bibinfo{year}{2009}\natexlab{}.
\newblock \bibinfo{booktitle}{{\em Search User Interfaces}}.
\newblock \bibinfo{publisher}{Cambridge University Press}.
\newblock


\bibitem[\protect\citeauthoryear{Huang, Ferraro, Mostafazadeh, Misra, Agrawal,
  Devlin, Girshick, He, Kohli, Batra, Zitnick, Parikh, Vanderwende, Galley, and
  Mitchell}{Huang et~al\mbox{.}}{2016}]%
        {DBLP:conf/naacl/HuangFMMADGHKBZ16}
\bibfield{author}{\bibinfo{person}{Ting{-}Hao~(Kenneth) Huang},
  \bibinfo{person}{Francis Ferraro}, \bibinfo{person}{Nasrin Mostafazadeh},
  \bibinfo{person}{Ishan Misra}, \bibinfo{person}{Aishwarya Agrawal},
  \bibinfo{person}{Jacob Devlin}, \bibinfo{person}{Ross~B. Girshick},
  \bibinfo{person}{Xiaodong He}, \bibinfo{person}{Pushmeet Kohli},
  \bibinfo{person}{Dhruv Batra}, \bibinfo{person}{C.~Lawrence Zitnick},
  \bibinfo{person}{Devi Parikh}, \bibinfo{person}{Lucy Vanderwende},
  \bibinfo{person}{Michel Galley}, {and} \bibinfo{person}{Margaret Mitchell}.}
  \bibinfo{year}{2016}\natexlab{}.
\newblock \showarticletitle{Visual Storytelling}. In \bibinfo{booktitle}{{\em
  {NAACL} {HLT} 2016, The 2016 Conference of the North American Chapter of the
  Association for Computational Linguistics: Human Language Technologies, San
  Diego California, USA, June 12-17, 2016}}. \bibinfo{pages}{1233--1239}.
\newblock
\showURL{%
\url{http://aclweb.org/anthology/N/N16/N16-1147.pdf}}


\bibitem[\protect\citeauthoryear{Kenter, Borisov, Gysel, Dehghani, de~Rijke,
  and Mitra}{Kenter et~al\mbox{.}}{2017}]%
        {DBLP:conf/sigir/KenterBGDRM17}
\bibfield{author}{\bibinfo{person}{Tom Kenter}, \bibinfo{person}{Alexey
  Borisov}, \bibinfo{person}{Christophe~Van Gysel}, \bibinfo{person}{Mostafa
  Dehghani}, \bibinfo{person}{Maarten de Rijke}, {and} \bibinfo{person}{Bhaskar
  Mitra}.} \bibinfo{year}{2017}\natexlab{}.
\newblock \showarticletitle{Neural Networks for Information Retrieval}. In
  \bibinfo{booktitle}{{\em Proceedings of the 40th International {ACM} {SIGIR}
  Conference on Research and Development in Information Retrieval, Shinjuku,
  Tokyo, Japan, August 7-11, 2017}}. \bibinfo{pages}{1403--1406}.
\newblock
\showDOI{%
\url{https://doi.org/10.1145/3077136.3082062}}


\bibitem[\protect\citeauthoryear{Kiseleva and de~Rijke}{Kiseleva and
  de~Rijke}{2017}]%
        {DBLP:journals/corr/KiselevaR17}
\bibfield{author}{\bibinfo{person}{Julia Kiseleva} {and}
  \bibinfo{person}{Maarten de Rijke}.} \bibinfo{year}{2017}\natexlab{}.
\newblock \showarticletitle{Evaluating Personal Assistants on Mobile devices}.
\newblock \bibinfo{journal}{{\em CoRR\/}}  \bibinfo{volume}{abs/1706.04524}
  (\bibinfo{year}{2017}).
\newblock
\showURL{%
\url{http://arxiv.org/abs/1706.04524}}


\bibitem[\protect\citeauthoryear{Li}{Li}{2015}]%
        {DBLP:phd/basesearch/Li15}
\bibfield{author}{\bibinfo{person}{Boyang Li}.}
  \bibinfo{year}{2015}\natexlab{}.
\newblock {\em \bibinfo{title}{Learning knowledge to support domain-independent
  narrative intelligence}}.
\newblock \bibinfo{thesistype}{Ph.D. Dissertation}. \bibinfo{school}{Georgia
  Institute of Technology, Atlanta, GA, {USA}}.
\newblock
\showURL{%
\url{http://hdl.handle.net/1853/53376}}


\bibitem[\protect\citeauthoryear{Li, Schijvenaars, and de~Rijke}{Li
  et~al\mbox{.}}{2017}]%
        {li-investigating-2017}
\bibfield{author}{\bibinfo{person}{Xinyi Li}, \bibinfo{person}{Bob
  Schijvenaars}, {and} \bibinfo{person}{Maarten de Rijke}.}
  \bibinfo{year}{2017}\natexlab{}.
\newblock \showarticletitle{Investigating queries and search failures in
  academic search}.
\newblock \bibinfo{journal}{{\em Information Processing \& Management\/}}
  \bibinfo{volume}{53}, \bibinfo{number}{3} (\bibinfo{date}{May}
  \bibinfo{year}{2017}), \bibinfo{pages}{666--683}.
\newblock


\bibitem[\protect\citeauthoryear{Marchionini}{Marchionini}{2006}]%
        {Marchionini:2006:Exploratory}
\bibfield{author}{\bibinfo{person}{Gary Marchionini}.}
  \bibinfo{year}{2006}\natexlab{}.
\newblock \showarticletitle{Exploratory Search: From Finding to Understanding}.
\newblock \bibinfo{journal}{{\em Commun. ACM\/}} \bibinfo{volume}{49},
  \bibinfo{number}{4} (\bibinfo{date}{April} \bibinfo{year}{2006}),
  \bibinfo{pages}{41--46}.
\newblock
\showISSN{0001-0782}
\showDOI{%
\url{https://doi.org/10.1145/1121949.1121979}}


\bibitem[\protect\citeauthoryear{Martin, Ammanabrolu, Hancock, Singh, Harrison,
  and Riedl}{Martin et~al\mbox{.}}{2017}]%
        {DBLP:journals/corr/MartinAHSHR17}
\bibfield{author}{\bibinfo{person}{Lara~J. Martin}, \bibinfo{person}{Prithviraj
  Ammanabrolu}, \bibinfo{person}{William Hancock}, \bibinfo{person}{Shruti
  Singh}, \bibinfo{person}{Brent Harrison}, {and} \bibinfo{person}{Mark~O.
  Riedl}.} \bibinfo{year}{2017}\natexlab{}.
\newblock \showarticletitle{Event Representations for Automated Story
  Generation with Deep Neural Nets}.
\newblock \bibinfo{journal}{{\em CoRR\/}}  \bibinfo{volume}{abs/1706.01331}
  (\bibinfo{year}{2017}).
\newblock
\showURL{%
\url{http://arxiv.org/abs/1706.01331}}


\bibitem[\protect\citeauthoryear{McIntyre and Lapata}{McIntyre and
  Lapata}{2010}]%
        {DBLP:conf/acl/McIntyreL10}
\bibfield{author}{\bibinfo{person}{Neil~Duncan McIntyre} {and}
  \bibinfo{person}{Mirella Lapata}.} \bibinfo{year}{2010}\natexlab{}.
\newblock \showarticletitle{Plot Induction and Evolutionary Search for Story
  Generation}. In \bibinfo{booktitle}{{\em {ACL} 2010, Proceedings of the 48th
  Annual Meeting of the Association for Computational Linguistics, July 11-16,
  2010, Uppsala, Sweden}}. \bibinfo{pages}{1562--1572}.
\newblock
\showURL{%
\url{http://www.aclweb.org/anthology/P10-1158}}


\bibitem[\protect\citeauthoryear{Miller, Feng, Fisch, Lu, Batra, Bordes,
  Parikh, and Weston}{Miller et~al\mbox{.}}{2017}]%
        {DBLP:journals/corr/MillerFFLBBPW17}
\bibfield{author}{\bibinfo{person}{Alexander~H. Miller}, \bibinfo{person}{Will
  Feng}, \bibinfo{person}{Adam Fisch}, \bibinfo{person}{Jiasen Lu},
  \bibinfo{person}{Dhruv Batra}, \bibinfo{person}{Antoine Bordes},
  \bibinfo{person}{Devi Parikh}, {and} \bibinfo{person}{Jason Weston}.}
  \bibinfo{year}{2017}\natexlab{}.
\newblock \showarticletitle{ParlAI: {A} Dialog Research Software Platform}.
\newblock \bibinfo{journal}{{\em CoRR\/}}  \bibinfo{volume}{abs/1705.06476}
  (\bibinfo{year}{2017}).
\newblock
\showURL{%
\url{http://arxiv.org/abs/1705.06476}}


\bibitem[\protect\citeauthoryear{Mrksic, S{\'{e}}aghdha, Wen, Thomson, and
  Young}{Mrksic et~al\mbox{.}}{2017}]%
        {DBLP:conf/acl/MrksicSWTY17}
\bibfield{author}{\bibinfo{person}{Nikola Mrksic},
  \bibinfo{person}{Diarmuid~{\'{O}} S{\'{e}}aghdha},
  \bibinfo{person}{Tsung{-}Hsien Wen}, \bibinfo{person}{Blaise Thomson}, {and}
  \bibinfo{person}{Steve~J. Young}.} \bibinfo{year}{2017}\natexlab{}.
\newblock \showarticletitle{Neural Belief Tracker: Data-Driven Dialogue State
  Tracking}. In \bibinfo{booktitle}{{\em Proceedings of the 55th Annual Meeting
  of the Association for Computational Linguistics, {ACL} 2017, Vancouver,
  Canada, July 30 - August 4, Volume 1: Long Papers}}.
  \bibinfo{pages}{1777--1788}.
\newblock
\showDOI{%
\url{https://doi.org/10.18653/v1/P17-1163}}


\bibitem[\protect\citeauthoryear{Munigala, Tamilselvam, and Sankaran}{Munigala
  et~al\mbox{.}}{2017}]%
        {creativity}
\bibfield{author}{\bibinfo{person}{Vitobha Munigala}, \bibinfo{person}{Srikanth
  Tamilselvam}, {and} \bibinfo{person}{Anush Sankaran}.}
  \bibinfo{year}{2017}\natexlab{}.
\newblock \showarticletitle{"Let me convince you to buy my product ... "}. In
  \bibinfo{booktitle}{{\em Proceedings of Workshop on Machine Learning for
  Creativity, SIGKDD, Nova Scotia, Canada, August 2017 (ML4Creativity 17)}}.
\newblock


\bibitem[\protect\citeauthoryear{Neumaier, Savenkov, and Vakulenko}{Neumaier
  et~al\mbox{.}}{2017a}]%
        {DBLP:journals/corr/NeumaierSV17}
\bibfield{author}{\bibinfo{person}{Sebastian Neumaier}, \bibinfo{person}{Vadim
  Savenkov}, {and} \bibinfo{person}{Svitlana Vakulenko}.}
  \bibinfo{year}{2017}\natexlab{a}.
\newblock \showarticletitle{Talking Open Data}.
\newblock \bibinfo{journal}{{\em CoRR\/}}  \bibinfo{volume}{abs/1705.00894}
  (\bibinfo{year}{2017}).
\newblock
\showURL{%
\url{http://arxiv.org/abs/1705.00894}}


\bibitem[\protect\citeauthoryear{Neumaier, Umbrich, and Polleres}{Neumaier
  et~al\mbox{.}}{2017b}]%
        {DBLP:conf/www/NeumaierUP17}
\bibfield{author}{\bibinfo{person}{Sebastian Neumaier},
  \bibinfo{person}{J{\"{u}}rgen Umbrich}, {and} \bibinfo{person}{Axel
  Polleres}.} \bibinfo{year}{2017}\natexlab{b}.
\newblock \showarticletitle{Lifting Data Portals to the Web of Data}. In
  \bibinfo{booktitle}{{\em Workshop on Linked Data on the Web co-located with
  26th International World Wide Web Conference {(WWW} 2017)}}.
\newblock
\showURL{%
\url{http://ceur-ws.org/Vol-1809/article-03.pdf}}


\bibitem[\protect\citeauthoryear{Radlinski and Craswell}{Radlinski and
  Craswell}{2017}]%
        {DBLP:conf/chiir/RadlinskiC17}
\bibfield{author}{\bibinfo{person}{Filip Radlinski} {and} \bibinfo{person}{Nick
  Craswell}.} \bibinfo{year}{2017}\natexlab{}.
\newblock \showarticletitle{A Theoretical Framework for Conversational Search}.
  In \bibinfo{booktitle}{{\em Proceedings of the 2017 Conference on Conference
  Human Information Interaction and Retrieval, {CHIIR} 2017, Oslo, Norway,
  March 7-11, 2017}}. \bibinfo{pages}{117--126}.
\newblock
\showDOI{%
\url{https://doi.org/10.1145/3020165.3020183}}


\bibitem[\protect\citeauthoryear{Riedl}{Riedl}{2016}]%
        {DBLP:journals/corr/Riedl16}
\bibfield{author}{\bibinfo{person}{Mark~O. Riedl}.}
  \bibinfo{year}{2016}\natexlab{}.
\newblock \showarticletitle{Computational Narrative Intelligence: {A}
  Human-Centered Goal for Artificial Intelligence}.
\newblock \bibinfo{journal}{{\em CoRR\/}}  \bibinfo{volume}{abs/1602.06484}
  (\bibinfo{year}{2016}).
\newblock
\showURL{%
\url{http://arxiv.org/abs/1602.06484}}


\bibitem[\protect\citeauthoryear{Thomas, McDuff, Czerwinski, and
  Craswell}{Thomas et~al\mbox{.}}{2017}]%
        {DBLP:journals/corr/Thomas17}
\bibfield{author}{\bibinfo{person}{Paul Thomas}, \bibinfo{person}{Daniel
  McDuff}, \bibinfo{person}{Mary Czerwinski}, {and} \bibinfo{person}{Nick
  Craswell}.} \bibinfo{year}{2017}\natexlab{}.
\newblock \showarticletitle{MISC: A data set of information-seeking
  conversations}.
\newblock  (\bibinfo{year}{2017}).
\newblock


\bibitem[\protect\citeauthoryear{Trippas, Spina, Cavedon, and
  Sanderson}{Trippas et~al\mbox{.}}{2017}]%
        {trippas2016how}
\bibfield{author}{\bibinfo{person}{Johanne~R Trippas}, \bibinfo{person}{Damiano
  Spina}, \bibinfo{person}{Lawrence Cavedon}, {and} \bibinfo{person}{Mark
  Sanderson}.} \bibinfo{year}{2017}\natexlab{}.
\newblock \showarticletitle{How Do People Interact in Conversational
  Speech-Only Search Tasks: A Preliminary Analysis}. In
  \bibinfo{booktitle}{{\em Proceedings of the 2017 ACM on Conference on Human
  Information Interaction and Retrieval (CHIIR)}}. ACM.
\newblock


\bibitem[\protect\citeauthoryear{White}{White}{2016}]%
        {white-interactions-2016}
\bibfield{author}{\bibinfo{person}{Ryen~W. White}.}
  \bibinfo{year}{2016}\natexlab{}.
\newblock \bibinfo{booktitle}{{\em Interactions with Search Systems}}.
\newblock \bibinfo{publisher}{Cambridge University Press}.
\newblock


\bibitem[\protect\citeauthoryear{White and Roth}{White and Roth}{2009}]%
        {DBLP:series/synthesis/2009White}
\bibfield{author}{\bibinfo{person}{Ryen~W. White} {and}
  \bibinfo{person}{Resa~A. Roth}.} \bibinfo{year}{2009}\natexlab{}.
\newblock \bibinfo{booktitle}{{\em Exploratory Search: Beyond the
  Query-Response Paradigm}}.
\newblock \bibinfo{publisher}{Morgan {\&} Claypool Publishers}.
\newblock
\showDOI{%
\url{https://doi.org/10.2200/S00174ED1V01Y200901ICR003}}


\bibitem[\protect\citeauthoryear{Williams, Raux, and Henderson}{Williams
  et~al\mbox{.}}{2016}]%
        {DBLP:journals/dad/WilliamsRH16a}
\bibfield{author}{\bibinfo{person}{Jason~D. Williams}, \bibinfo{person}{Antoine
  Raux}, {and} \bibinfo{person}{Matthew Henderson}.}
  \bibinfo{year}{2016}\natexlab{}.
\newblock \showarticletitle{The Dialog State Tracking Challenge Series: {A}
  Review}.
\newblock \bibinfo{journal}{{\em D{\&}D\/}} \bibinfo{volume}{7},
  \bibinfo{number}{3} (\bibinfo{year}{2016}), \bibinfo{pages}{4--33}.
\newblock
\showURL{%
\url{http://dad.uni-bielefeld.de/index.php/dad/article/view/3685}}


\bibitem[\protect\citeauthoryear{Zhang, S{\'{e}}aghdha, Quercia, and
  Jambor}{Zhang et~al\mbox{.}}{2012}]%
        {DBLP:conf/wsdm/ZhangSQJ12}
\bibfield{author}{\bibinfo{person}{Yuan~Cao Zhang},
  \bibinfo{person}{Diarmuid~{\'{O}} S{\'{e}}aghdha}, \bibinfo{person}{Daniele
  Quercia}, {and} \bibinfo{person}{Tamas Jambor}.}
  \bibinfo{year}{2012}\natexlab{}.
\newblock \showarticletitle{Auralist: introducing serendipity into music
  recommendation}. In \bibinfo{booktitle}{{\em Proceedings of the Fifth
  International Conference on Web Search and Web Data Mining, {WSDM} 2012,
  Seattle, WA, USA, February 8-12, 2012}}. \bibinfo{pages}{13--22}.
\newblock
\showDOI{%
\url{https://doi.org/10.1145/2124295.2124300}}


\end{thebibliography}

\balance

\end{document}